       \documentclass[12pt,preprint]{aastex}
\newcommand{\myemail}{D.TRIPATHI@damtp.cam.ac.uk}
\shorttitle{Sigmoid temperature tomography and flux rope}
\shortauthors{Tripathi et al.}
\begin{document}
%%----------------------------------------------------------------------------------------------
\title{Temperature Tomography of a Coronal Sigmoid Supporting the Gradual Formation of a Flux Rope}
\author{Durgesh Tripathi\altaffilmark{1}, 
        Bernhard Kliem\altaffilmark{2,3,4}, 
        Helen E. Mason\altaffilmark{1}, 
        Peter R. Young\altaffilmark{3,5}, and
        % J. Len Culhane\altaffilmark{2}, and
        Lucie M. Green\altaffilmark{2}}
\affil{$^1$DAMTP, University of Cambridge, Wilberforce Road, Cambridge
           CB3 0WA, UK}
\affil{$^2$University College London, Mullard Space Science Laboratory,
           Holmbury St.\ Mary, Dorking, Surrey, RH5 6NT, UK}
\affil{$^3$Space Science Division, Naval Research Laboratory, 
           Washington, DC 20375, USA}
\affil{$^4$Institut f{\"u}r Physik und Astronomie, Universit{\"a}t Potsdam,
           Potsdam 14482, Germany}
\affil{$^5$George Mason University, 4400 University Drive, Fairfax, VA 22030, USA}           
\email{\myemail}

\date{\today}

%----------------------------------------------------------------------------------------------
\begin{abstract} 
Multi-wavelength observations of a sigmoidal (S-shaped) solar coronal source by the EUV Imaging Spectrometer and the X-ray Telescope aboard the \textsl{Hinode} spacecraft and by the EUV Imager aboard \textsl{STEREO} are reported. The data reveal the coexistence of a pair of J-shaped hot arcs at temperatures $T>2$~MK with an S-shaped structure at somewhat lower temperatures ($T\approx1\mbox{--}1.3$~MK). The middle section of the S-shaped structure runs along the polarity inversion line of the photospheric field, bridging the gap between the arcs. Flux cancellation occurs at the same location in the photosphere. The sigmoid forms in the gradual decay phase of the active region, which does not experience an eruption. These findings correspond to the expected signatures of a flux rope forming, or being augmented, gradually by a topology transformation inside a magnetic arcade. In such a transformation, the plasma on newly formed helical field lines in the outer flux shell of the rope (S-shaped in projection) is expected to enter a cooling phase once the reconnection of their parent field line pairs (double-J shaped in projection) is complete. Thus, the data support the conjecture that flux ropes can exist in the corona prior to eruptive activity.
\end{abstract}
%%----------------------------------------------------------------------------------------------
%---
\section{Introduction}\label{sec:Introduction}
%---

The basic magnetic topology of active regions is a key factor in determining their potential for producing an eruption and is a subject of intense debate. It is accepted that soft X-ray and EUV loops illuminate field lines with an arcade-like topology overlying a polarity inversion line (PIL) of the photospheric field. There is also substantial evidence that erupting filaments/prominences possess a flux rope topology once they have risen somewhat from their original position. This is strongly indicated by H$\alpha$, EUV, and coronagraph images \cite[e.g.,][]{Romano&al2003} and supported by successful fitting of flux rope models to interplanetary in situ data \citep{Jian&al2006}. However, the question remains elusive as to when and how the flux rope forms. The following scenarios have been suggested.
\newline\hspace*{\parindent}
1) The region's birth as a flux rope by bodily emergence of the parent rope in the convection zone \cite[e.g.,][]{Rust&Kumar1994, Low1996, Gibson&al2004}.
\newline\hspace*{\parindent}
2) Formation immediately after the emergence of the active region by multiple reconnections near the PIL, driven by the shear flows associated with the emergence \citep{Manchester&al2004, Archontis&Torok2008}.
\newline\hspace*{\parindent}
3) Gradual formation between emergence and eruption by multiple reconnections near the PIL, due to flows converging at the PIL and magnetic diffusion in the photosphere \cite[e.g.,][]{Amari&al2003a, Amari&al2003b}.
\newline\hspace*{\parindent}
4) Formation immediately before and in the course of the eruption by initially slow, then accelerating reconnection in the corona above the PIL \citep{Moore&al2001}.
\newline\hspace*{\parindent}
5) Formation only after the onset of an eruption through fast reconnection in the corona triggered by the eruption \citep{MacNeice&al2004}.

In cases (2--5) the flux rope is formed by a topological transformation in the inner arcade \citep{vanBallegooijen&Martens1989}. However, the requirement of nearly force-free equilibrium in cases (1-4) implies that the rope remains embedded in an arcade \cite[e.g.,][]{Titov&Demoulin1999}, i.e., the outer part of the arcade is not transformed into a flux rope unless the configuration erupts. We are primarily interested in the question whether the flux rope topology can be formed prior to eruptions, as this is a key issue for their theoretical modeling. The gradual flux rope formation, as suggested in case (3), appears to be most relevant in this regard, because eruptions typically do not develop in close temporal association with the emergence of active regions, but rather tend to occur delayed by several days.

The gradual transformation of magnetic topology is thought to be driven by photospheric flows and diffusion of the magnetic field in the vicinity of the PIL, with converging flows being most efficient \cite[e.g.,][]{Inhester&al1992, Amari&al2003a, Amari&al2003b}. Both have the observational signature of so-called flux cancellation in the photosphere. However, this is a quite common process \cite[see, e.g.,][]{tripathi05}, which can also result from a simple submergence of loops. Therefore, flux cancellation cannot be considered as a sufficient indicator of flux rope formation.

All of the suggested flux rope formation mechanisms involve magnetic reconnection at or above the PIL. An obvious signature of this process would be the emission from the resulting hot plasma. Good candidates for such emission are sigmoidal (S or reverse-S shaped) X-ray sources because their shape is indicative of twisted field. Moreover, their occurrence is correlated with that of an eruption \citep{Canfield&al1999}. Sigmoids exhibit a transient brightening when the region erupts, which additionally points to a connection between such sources and the eruption process, which may be driven by a flux rope instability \cite[e.g.,][]{Kliem&Torok2006}. Unambiguous signatures of reconnection for a gradually forming flux rope have been hard to observe \cite[e.g.,][]{McKenzie&Canfield2008}, presumably because such reconnection generally has a complex (intermittent and patchy) occurrence pattern, forming a rope in many small steps (except in case 5). Only very recently, \citet{Green&Kliem2009} identified a sigmoidal active region in which a flux rope formed through reconnection over a period of at least several hours.

The structure and dynamics of filaments/prominences, which trace out part of the field topology in the stable as well as in the eruptive phases of the region's evolution, is potentially the best available source of information about the topology. However, it is still considered to be an open question whether the observations and the modeling of these structures in their stable phase favor the arcade topology \cite[e.g.,][]{Martin1998, Wang2001} or the flux rope topology \cite[e.g.][]{Rust&Kumar1994, Aulanier&al2000, Bobra&al2008}.

Here, we pursue a novel approach to identify observational signatures of the presence of a coronal magnetic flux rope, which relies on the efficient thermal insulation by the magnetic field in the nearly collisionless coronal plasma. At any instant during the topology transformation from an arcade to a flux rope, reconnection takes place only at the rope's outermost flux shell which is just being formed. The previously reconnected flux shells detach from the volume of reconnection where the hottest plasma is produced. Being the result of reconnection of two ordinary loop field lines near their closest footpoints, the new flux rope field lines are much longer and carry more plasma than each of the original loops. For these reasons, the volumetric heating rate is likely reduced after the reconnection is complete and the plasma in the newly formed flux shells is expected to enter a cooling phase. According to this logic, if a filament/prominence forms, it is likely to reside near the magnetic axis of the rope. However, this is the place where the rope topology is least apparent, since the axis tends to be straighter than the field lines in the outer flux shells that wind about the axis. Therefore, observing plasma that has just started to cool from X-ray emitting temperatures has the highest potential to reveal a flux rope topology. Such cooling can be observed best for a gradually forming flux rope.

The EUV Imaging Spectrometer \citep[EIS;][]{eis} aboard the \textsl{Hinode} spacecraft \citep{hinode} is the ideal instrument for carrying out such observations, as it yields images in individual spectral lines formed in the required range of temperatures. Previous multi-wavelength sigmoid investigations using the CDS aboard the \textsl{SoHO} satellite by \cite{Gibson&al1999, Gibson&al2002} and \cite{tripathi06} may not have had sufficient spatial resolution to discover the effect we are studying here. In the following sections we analyze the first observation of a sigmoid recorded with EIS, together with soft X-ray, EUV filter imaging, and photospheric magnetic field data. 

%---
\section{Observations and Data Analysis}\label{obs}
%---

%%----------------
\begin{figure}
	\centering
	\includegraphics[width=0.8\textwidth]{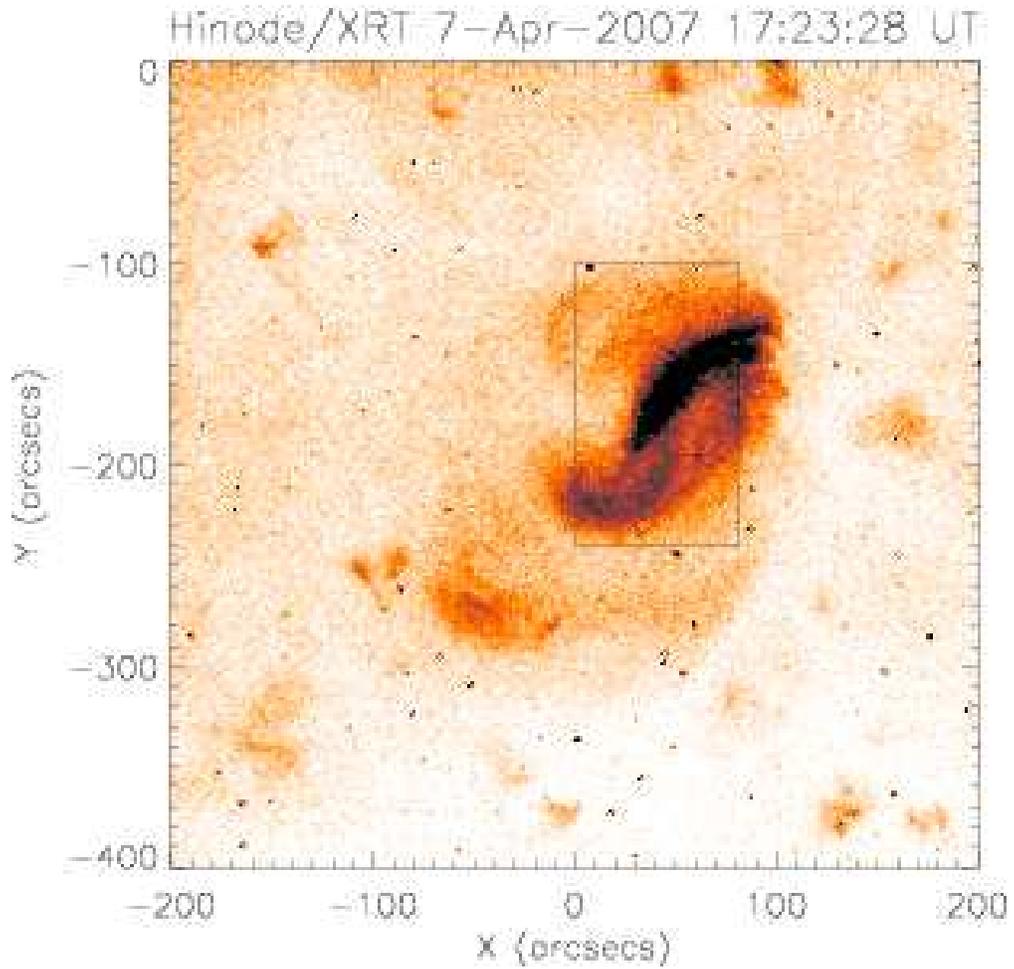}
	\caption{\textsl{Hinode} XRT image showing the sigmoidal region at the time when the EIS slit was located approximately in the middle of the raster. The EIS raster area displayed in Fig.~\ref{eis} is indicated by the box.\label{context}}
\end{figure}
%%----------------

The EIS instrument resolves the spectral ranges 170--210 and 250--290~{\AA}, providing observations in a broad range of temperatures ($\log T[\mathrm{K}]=4.7\mbox{--}7.3$). The data analyzed here were obtained from a single raster that scanned a field of view (FOV) of 100{\arcsec} by 216{\arcsec} with the 2{\arcsec} slit from west to east with an exposure time of 30~sec. The count rates permitted us to assemble images from extracted spectral lines that are formed in the temperature range $\log T=5.8\mbox{--}6.4$. 

Filter images by the EUV Imager (EUVI) aboard the \textsl{STEREO-A} satellite \citep{Howard&al2008} in the \ion{Fe}{12} 195~{\AA} and \ion{Fe}{15} 284~{\AA} lines yield complementary information about the temporal evolution of the sigmoid at the corresponding temperatures of $\log T=6.1$ and 6.4, respectively.

The X-Ray Telescope \citep[XRT;][]{xrt} aboard \textsl{Hinode} provides partial-frame coronal images at high cadence and full-disk images every 6 hours with 1{\arcsec} pixels. The latter are useful for the co-alignment with images from other instruments. The sigmoidal region was observed at a cadence of $\sim$1~min with the \textit{Al\_poly} filter, which selects plasma in a wide range of temperatures with peak sensitivity at $\log{T}=6.9$.

There are no corresponding photospheric data available from the optical telescope on \textsl{Hinode}. Therefore, we use line-of-sight magnetograms recorded by the Michelson Doppler Imager \citep[MDI;][]{mdi} on \textsl{SoHO}.

We have applied the standard data reduction software for EIS, EUVI, XRT, and MDI data as provided by the Solar Software (SSW) package. Full-disk MDI and XRT images were co-aligned by limb fitting. The XRT partial-frame images were then co-aligned with full-disk images by cross-correlating specific features. The \ion{Fe}{15} raster image was used for the EIS-XRT co-alignment through cross-correlation, as this image is closest in temperature and appearance to the XRT images. We believe that the co-alignment thus achieved is quite accurate, with an error of about 2--5~arcsec. Our study does not require the complex task of coaligning the EIS and EUVI images.

The target region was selected because of its sigmoidal appearance and location near Sun center on the day of observation, 7 April 2007. It was a weak, spot-less, old active region, which showed a substantial dispersal of flux already as it appeared at the east limb. It entered its final decay after 8 April. The region did not have a filament and neither flared nor produced a mass ejection. A sudden change of the diffuse outer loops of the region visible in EUVI \ion{Fe}{15} images between 6 April 23:40~UT and 7 April 00:40~UT, did not affect the bright, highly sheared loops in the region's core. Figure~\ref{context} displays an XRT image of the sigmoidal region at the time of the EIS raster scan, providing context information. The XRT observed the region in partial-frame mode from 10:22:06 to 17:58:03~UT, which includes the FOV and time of the EIS raster (17:09:32--17:35:49~UT).

%%-------------
\section{Results and discussion} \label{results}
%%-------------

%%-----------------
\begin{figure*}
	\centering
	\includegraphics[width=0.8\textwidth]{xrt_figure.eps}
	\caption{XRT images showing the evolution of the two J-shaped structures toward a continuous sigmoid (marked by the arrows). The images were differentially rotated to the time when the EIS slit was located in the middle of the raster ($\approx$~17:24~UT).  An animation showing the full evolution is available online.\label{xrt}}
\end{figure*}
%%-----------------
%%-----------------
\begin{figure*}
	\centering
	\includegraphics[width=0.8\textwidth]{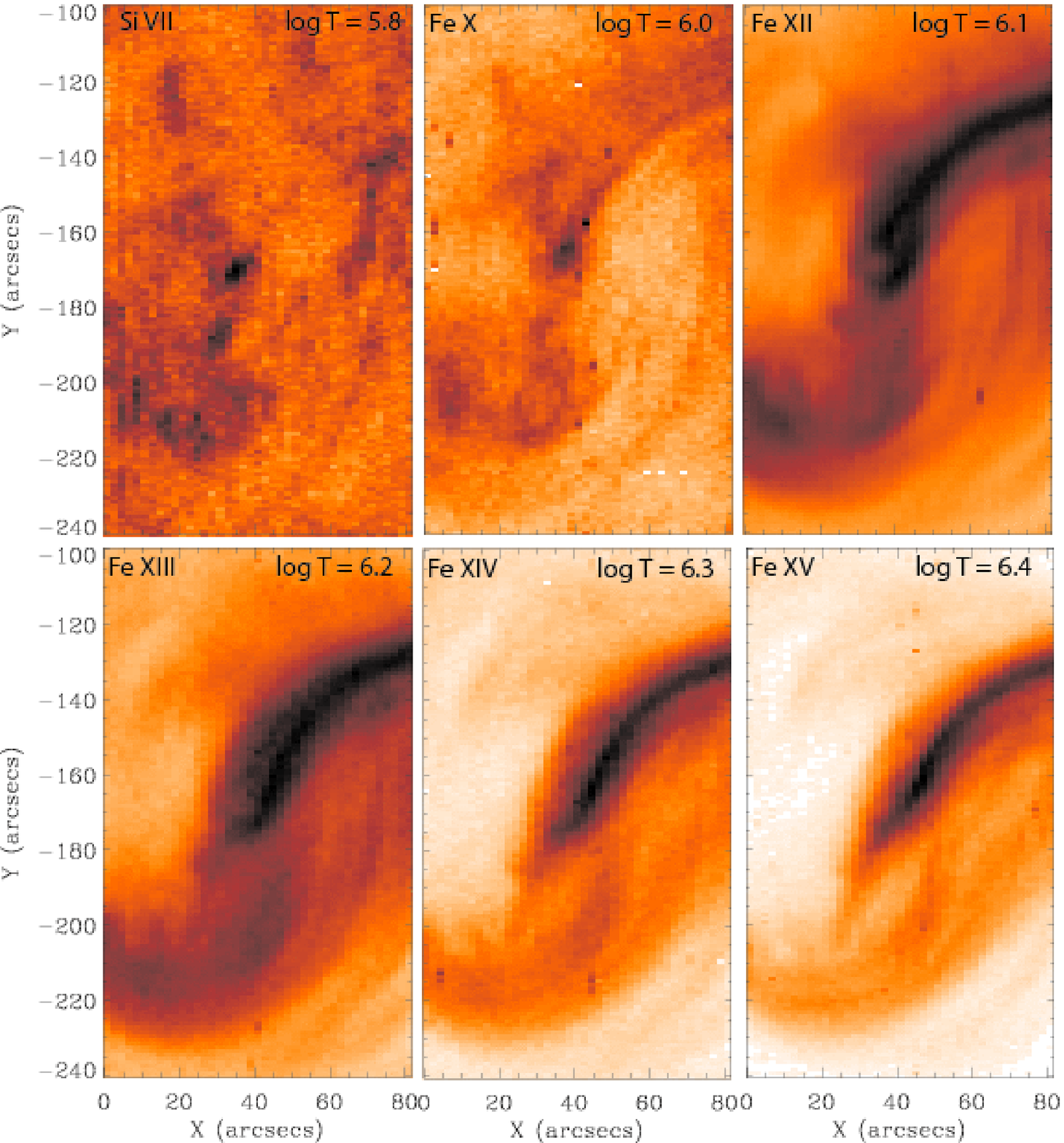}
	\caption{EIS intensity maps in different emission lines obtained simultaneously from the raster scan between 17:09:32 and 17:35:49~UT.\label{eis}}
\end{figure*}
%%----------------

Figure~\ref{xrt} displays XRT images showing the evolution of the hot coronal sources. The region is comprised of two J-shaped arcs (each consisting of multiple threads, or loops), labelled J1 and J2. In the course of the observation, the lower (southern) arc J2 splits in two parts, labelled A and B. The upper part, A, starts to diffuse away subsequently, and the lower part, B, moves eastward. The upper footpoint of the feature B, marked by an arrow in the two right panels, approaches the lower footpoint of the upper (northern) arc J1 (see the bottom right image), as if the two J-shaped structures J1 and B are in the process of merging to form a single S-shaped structure. 

Figure~\ref{eis} displays images simultaneously obtained in different spectral lines from the EIS raster scan. The images obtained in \ion{Fe}{14} and \ion{Fe}{15} resemble the XRT images best, as expected from their formation temperatures of $\log{T}=(6.3\mbox{--}6.4)$. They exhibit the two J-shaped arcs clearly. In contrast, the \ion{Si}{7} image shows very little correspondence to the J-shaped structures seen by XRT, which clearly demonstrates that the sigmoid has only very minor contribution from plasma at the formation temperature of this relatively bright line, $\log{T}=5.8$, suggesting an ongoing heating in the sigmoid.

\begin{figure*}
	\centering
	\includegraphics[width=0.8\textwidth]{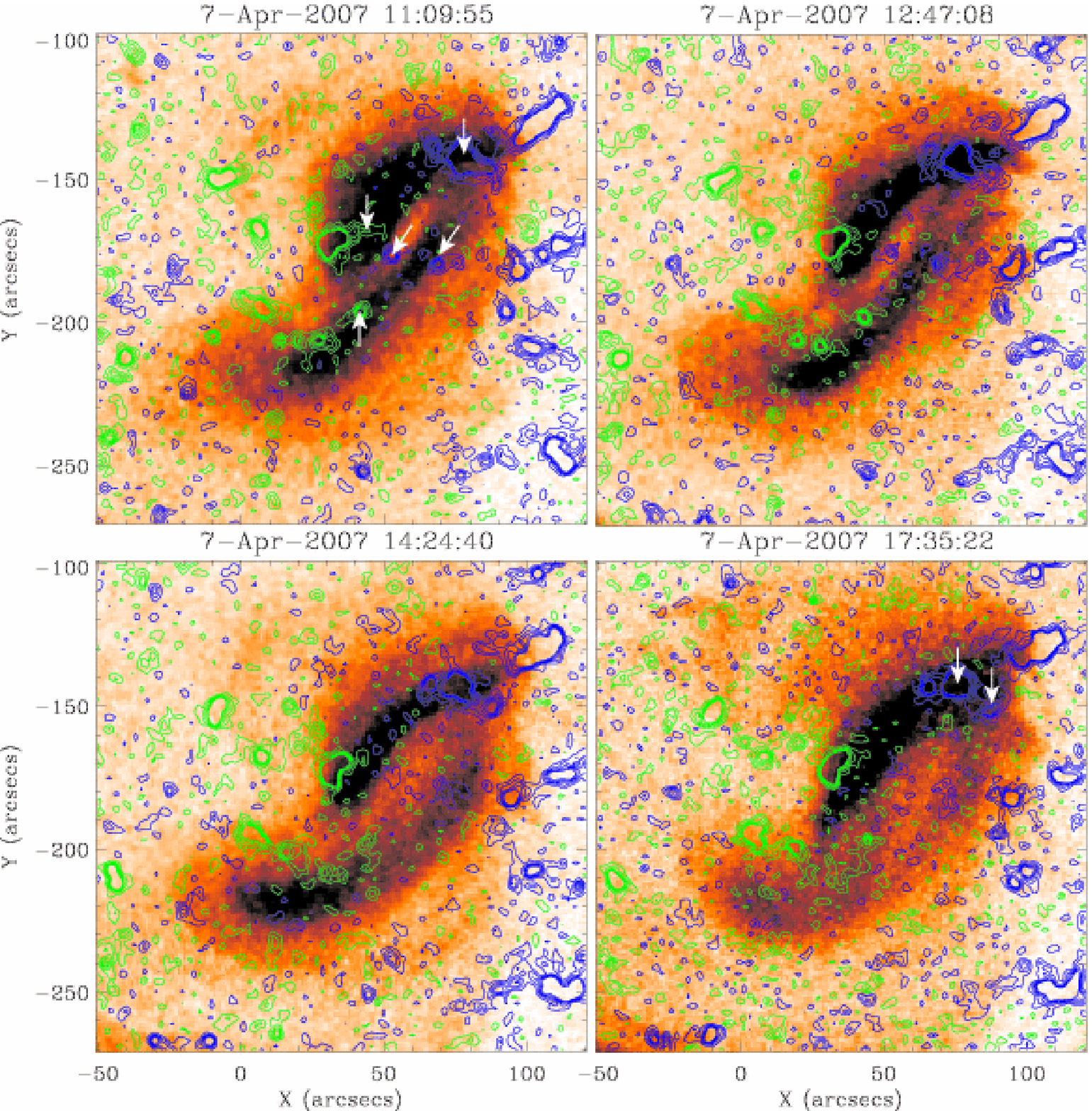}
	\caption{\label{xrt_mdi}
	MDI magnetograms (contours) overplotted on partial-frame XRT images corresponding closest in time. Blue (green) contours represent positive (negative) polarity. All images are differentially rotated to the time when the EIS slit was located in the middle of the raster ($\approx$~17:24~UT). The arrow pointing at $(x,y)=(70,-140)$ in the first panel marks the positive patch of flux that moves inward and away from the other positive flux at the end of the arc J1, eventually even splitting in two patches (see arrows in the fourth panel). The other four arrows in the first panel mark the flux elements that cancel in the course of the day (note that the major negative flux patch near the southern end of arc J1 survives, but its marked extension, which was a separate patch in the beginning of the day, cancels.) The magnetic evolution during the whole day is shown in a corresponding animation.}
\end{figure*}

The images obtained at intermediate temperatures in \ion{Fe}{10} and \ion{Fe}{12} ($\log{T}=6.0\mbox{--}6.1$) do not show two separate J's as the dominant sources. Instead they display an S-shaped structure bridging the gap between the two J's. The S is not of uniform brightness. We interpret this as resulting from a superposition of a diffuse and relatively weak but continuous S-shaped source with background loops that connect some of the numerous small flux patches in the area (see below and Figure~\ref{xrt_mdi}). The patchy appearance of the sigmoid in the \ion{Fe}{10} image is primarily due to the weakness of this line in comparison to \ion{Fe}{12}. Finally, the image obtained in \ion{Fe}{13} ($\log{T}=6.2$) shows a mixture of the double J and the S.

%%----------------
\begin{figure*}
	\centering
	\includegraphics[width=0.8\textwidth]{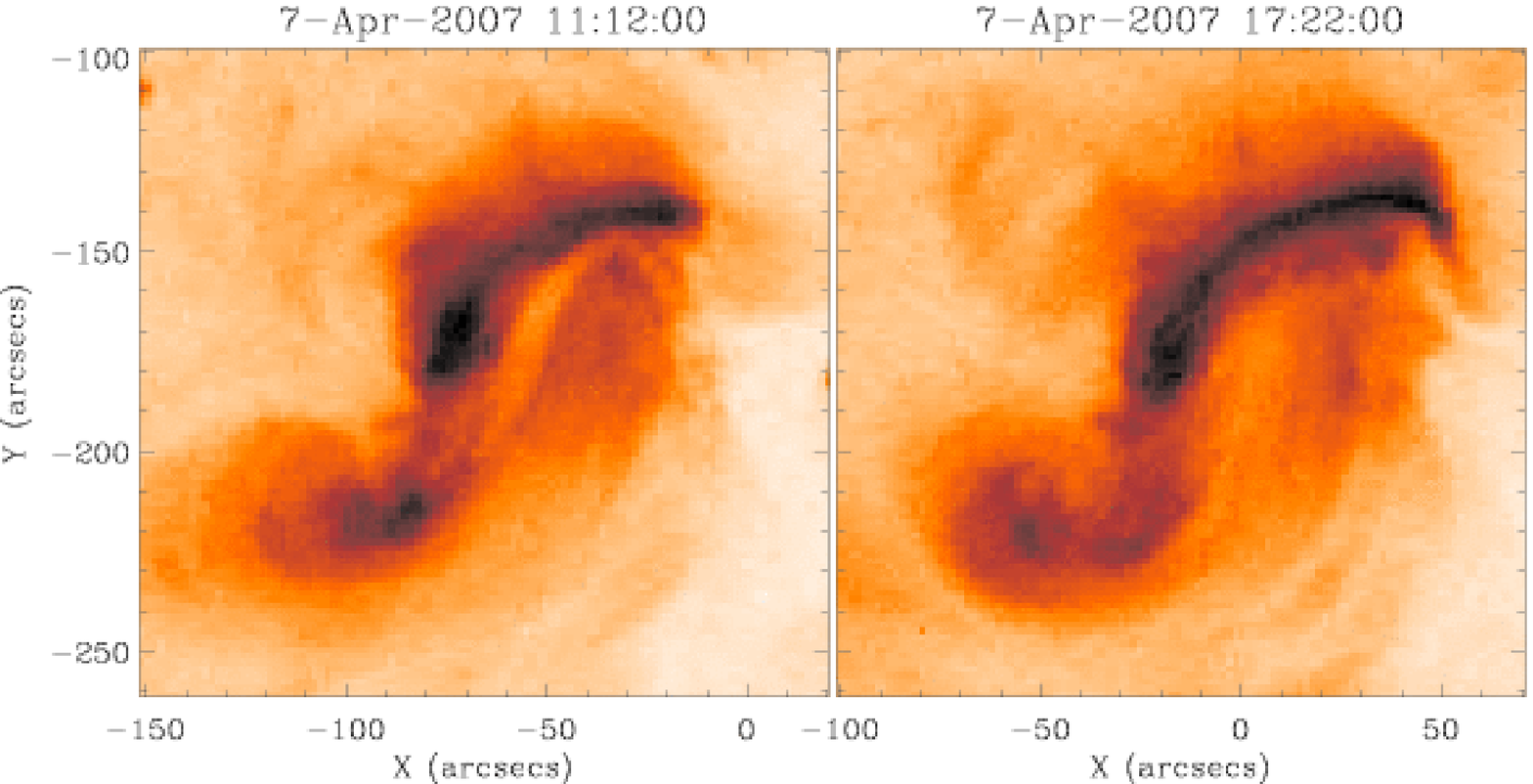}
	\caption{\label{euvi}
	EUVI 195~{\AA} images of the sigmoid at the beginning of the XRT observation (\textit{left}) and at a time in the middle of the EIS raster scan (\textit{right}).}
\end{figure*}
%%----------------

The co-existence of the double-J source at high temperatures with the S-shaped source at adjacent lower temperatures is in agreement with the expectations for a flux rope forming inside an arcade by reconnection near the PIL (Section~\ref{sec:Introduction}). In this picture, the reconnection occurs in the outermost flux shell of the forming flux rope in the vicinity of the PIL, producing hot plasma. This volume is threaded by arcade field lines with a split J shape (in photospheric projection) in the lateral reconnection inflow regions and by flux rope field lines with continuous S-shaped projection in the upward reconnection outflow region. Field lines of the first category are indicated by the two J-shaped arcs in the high temperature images ($\log{T}\ge6.2$) after $\approx$~16:30~UT (Figures~\ref{xrt} and \ref{eis}). The plasma entering the flux rope in the reconnection outflow starts to cool, as indicated by the appearance of the sigmoid in the \ion{Fe}{10} and \ion{Fe}{12} images.

Brightness in these images is not proportional to temperature, but rather to column density squared in the selected temperature range. Hence, the brightness of the cooling plasma in the outer flux shell of a forming rope depends on the temperature, amount of hot plasma on the newly reconnected field lines, and the thickness of the flux shell. Therefore, it may vary in a wide range. In the present case, since the active region is highly dispersed, the reconnection rate and, therefore, the rate at which hot plasma is added to newly reconnected field lines should be relatively small. Hence, the brightness may be rather low, consistent with the appearance of the sigmoid in \ion{Fe}{12}. The strong difference in brightness between the hot arcs J1 and J2 can be related to different amounts of magnetic flux threading them (as indicated by the MDI data), since the soft X-ray emission measure is proportional to the magnetic flux in the source \citep{Pevtsov&al2003, Tan&al2007}.

Figure~\ref{euvi} displays two EUVI \ion{Fe}{12} 195~{\AA} images taken at the beginning of the XRT observation and at the time of the EIS raster, the latter corresponding closely to the EIS \ion{Fe}{12} image. These images demonstrate that the S-shaped sigmoid exists at these temperatures for several hours prior to the EIS raster scan. The earliest time at which it can be identified is $\sim$~04:00~UT.

%%----------------
%%----------------

The evolution of the photospheric flux is related to the coronal evolution in Figure~\ref{xrt_mdi}. The fragmented nature of the flux distribution makes it difficult to obtain clear-cut associations. The three major flux concentrations in the area of the sigmoid are located near the ends of the upper X-ray arc J1 and near the far (lower) end of the lower X-ray arc J2. They are partially split in a series of fragments, and the positive flux concentration actually consists of two major patches. It is difficult to determine the exact location and shape of the PIL and how it changes, but clearly the PIL crosses the sigmoid in the middle, in the area where the S-shaped trace is observed in \ion{Fe}{10} and \ion{Fe}{12}, and roughly in north-south direction.

The flux fragments execute more or less irregular motions during the day, with two systematic changes standing out (see also the corresponding animation). The first is the cancellation of four flux patches in the middle of the sigmoid (marked by arrows in the first panel). The two closest of these patches determine the initial location of the PIL in this area rather sharply. The second systematic change is a splitting of the double positive flux patch near the northern ends of the arcs J1 and J2, with a southeastward motion (toward the PIL in the center of the sigmoid) of the eastern flux patch and subsequent disintegration into smaller patches (see arrows in the first and fourth panels). Some threads in the arc J2 appear to follow the southeastward moving flux patch, which suggests that the splitting of J2 into the components A and B is a consequence of these flux changes.

The location of the canceling flux patches shows that the PIL runs under the middle section of the S-shaped sigmoid in \ion{Fe}{10} and \ion{Fe}{12} and between the double J-shaped sigmoid in \ion{Fe}{14} and \ion{Fe}{15}, which is also where the X-ray structures B and J1 are approaching. Thus, although not providing conclusive evidence, the photospheric evolution does support the picture of gradual flux rope formation indicated by the multi-temperature coronal data.

A screening of the data back to 4 April\footnote{available at http://virtualsolar.org} reveals that the region, at times and beginning on 5 April, exhibited an arrangement of its loops that resembled a sigmoid. The flux cancellation seen on 7 April had already begun in the middle of the preceding day, accompanied by a faint, temporary J-shaped extension of the active region loops toward the southeast. Therefore, it is possible that XRT and EIS observed a phase of augmentation of a flux rope that formed over a longer time period, or that even existed since the region's emergence (albeit the latter does not find direct support in the data presented here).

After the EIS observation, the sigmoid gradually begins to disintegrate. For $\sim4$ hours the approach of the hot structures J1 and B, visible also in EUVI \ion{Fe}{15}, continues. However, B does not merge with J1 subsequently, but rather forms two thin S-shaped threads which extend parallel to J1 with slightly and irregularly varying positions. The continuous S in the EUVI \ion{Fe}{12} images shows a disintegration into similar threads, which remain in the spatial location between the threads and the arc J1 visible in \ion{Fe}{15}. After $\sim10$~UT on 8 April the eastern end of J1 moves apart from the center of the sigmoid, and by the end of the day the region appears no longer sigmoidal. After 8 April, a rapid final dissolution of the region is seen, both in the corona and photosphere, destroying the spatial coherence which is indicated by the sigmoid, thus counteracting the evolution toward an eruption.

%%-------------
\section{Summary and Conclusions} \label{con}
%%-------------

The present investigation of a coronal sigmoidal source in a range of X-ray and EUV emitting temperatures ($\log{T[K]}\sim5.8\mbox{--}6.9$) has revealed the coexistence of a double J-shaped and an S-shaped structure separated by temperature during a period of several \textbf{($>7$)} hours in a decaying active region. The S-shaped structure bridges the gap in the double-J source. 
This is consistent with the topology of a magnetic flux rope forming, or being augmented, gradually by reconnection inside an arcade supporting cases (1\mbox{--}3) in Section~\ref{sec:Introduction}. X-ray and magnetogram data reveal the occurrence of the combined structure to be embedded in an evolution that is also consistent with such topology transformation. Flux cancellation occurs where the S source connects the two J sources. Moreover, one leg of the double J-shaped X-ray source and one of the major flux concentrations in the magnetogram exhibit motions converging toward the PIL.

The data do not permit us to determine when the suggested flux rope formation commenced. Since the structure became apparent in the course of the region's gradual evolution, models of eruptions are supported that assume a coronal flux rope can be built up prior to the event.

While our results regarding the flux rope topology are of an indicative nature and require substantiation through further similar observations, the present study clearly emphasizes the importance of multi-wavelength observations in studying the evolution of source regions of eruptive events.

\acknowledgments

We thank the referee and S.~Patsourakos for very constructive comments. \textsl{Hinode} is a Japanese mission developed and launched by ISAS/JAXA, with NAOJ as domestic partner and NASA and STFC (UK) as international partners. It is operated by these agencies in co-operation with ESA and NSC (Norway). This work was supported by STFC Rolling Grants and by NASA grant NNH06AD58I.

%%----------------------------

\end{document}